\documentstyle[prb,floats,twocolumn,aps,epsf]{revtex}
\begin{document}
%\draft
\noindent{\Large{\it A posteriori} teleportation\hfill}

\vskip0.1truein
\noindent
{Samuel L.~Braunstein${}^\ast$ and H.\ J.\ Kimble${}^\dagger$}
\vskip0.1truein
\noindent
{${}^\ast$SEECS, University of Wales, Bangor LL57 1UT, UK\\
${}^\dagger$Norman Bridge Laboratory of Physics 12-33, \\
California Institute of Technology, Pasadena, CA 91125}
% \date{\today}
\vskip 0.05truein
\noindent
{\vrule width\columnwidth height0.4pt depth0pt\relax}

The article by Bouwmeester {\it et al}.\cite{bouwmeester97} on experimental 
quantum teleportation constitutes an important advance in the burgeoning 
field of quantum information. The experiment was motivated by the proposal 
of Bennett {\it et al}.\cite{bennett93} in which an unknown quantum state 
is `teleported' by Alice to Bob. As illustrated in Fig.~1, in the 
implementation of this procedure, by Bouwmeester {\it et al}., an input 
quantum state is `disembodied' into quantum and classical components, as 
in the original protocol.\cite{bennett93} However, in contrast to 
the original scheme, Bouwmeester {\it et al}.'s procedure necessarily 
destroys the state at Bob's receiving terminal, so a `teleported' state 
can never emerge as a freely propagating state for subsequent examination 
or exploitation. In fact, teleportation is achieved only as a postdiction.

Bouwmeester {\it et al}. used parametric down-conversion from two sources 
(SI, SII in Fig.~1) in an attempt to teleport the polarization state of 
a single-photon wavepacket (in beam 1) from Alice's sending station to 
Bob's receiving station (in beam 3). Statistics consistent with 
teleportation are obtained for events with a fourfold coincidence (from 
detectors f1 and f2, d1 or d2, and p). We ask whether the detection of all
four quanta is essential for teleportation in this scheme. To answer this 
question, we calculated the teleportation fidelity, $F$, when the coincidence 
condition is relaxed to exclude detection at Bob's station (d1, d2). 

Under relaxed conditions, requiring only threefold coincidence of detectors 
p and (f1, f2), teleportation is achieved when the fields of beams 1
and 3 match with sufficiently high fidelity. In the simplest
approximation, type II parametric down-conversion on modes ($i$, $j$)
generates wavepacket states as follows: 
\begin{equation}
A_{0}|0\rangle _{ij}+A_{1}|\psi ^{-}\rangle _{ij}+A_{2}|\chi \rangle
_{ij}+\cdots \;,  \label{psi}
\end{equation}
where $A_{0}$, $A_{1}$ and $A_{2}$ are the coefficients for obtaining no
(vacuum), one and two down-converted pairs, respectively, and 
($i$, $j$) = ($1$, $4$), ($2$, $3$). Of these terms, only states 
corresponding to the second term are selected by fourfold coincidence, 
as specified by equations~(2) and~(3) of ref.~\onlinecite{bouwmeester97}. 
However, anything less than complete destruction of the output 3 
necessarily leaves undesirable terms that reduce $F$.

% FIG 1
\begin{figure}[ht]
\epsfxsize=3.0in
\epsfbox[-20 140 580 610]{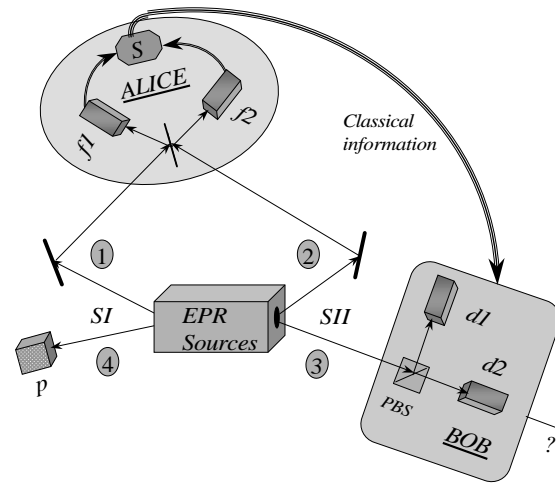}
\caption{The teleportation set-up of ref.~\protect\onlinecite{bouwmeester97}).
PBS, polarizing beam splitter.}
\label{fig1}
\end{figure}

The initial input state to Alice's station, $|\Phi\rangle$, is prepared by
detecting the state in field 4 at detector p, which projects the field in
beam 1 accordingly. Joint detection at ($f1$, $f2$) then provides
threefold coincidence with p, yielding a statistical mixture for the field 3
arriving at Bob's station. The fraction of the state $|\Phi\rangle$ in
this mixture gives $F$. To lowest order in the down-converter coupling
strength, the Bouwmeester {\it et al}.\ scheme yields a 50:50 mixture of the
vacuum state $|0\rangle $ and the desired state $|\Phi\rangle $, with 
$F=1/2$, so that there is never a physical state with high teleportation 
fidelity. Indeed, Bob could achieve this same fidelity, $F=1/2$, 
by abandoning teleportation altogether and transmitting randomly selected 
polarization states.  Faced with this state of affairs, the experiment of 
ref.~\onlinecite{bouwmeester97} obtains a surrogate for high fidelity by 
destructively recording the field 3 at (d1, d2).

We emphasize that the nature of the mixture containing the vacuum state
has definite physical implications, which can be verified by more general 
measurements than photon counting (for example, by quantum-state 
tomography). Moreover, the freedom of a potential consumer of the output 
from Bob's receiving station to select alternative detection strategies 
means that classical analogies fail.

To achieve conventional {\it a priori} teleportation, the setup in
ref.~\onlinecite{bouwmeester97} would have to be modified to eliminate 
the vacuum from the mixture. Because the vacuum appears when two pairs of 
(1, 4) photons are created, we might seek to resolve one- and two-photon 
detection events at p. Upgraded detection (for example, by cascading 
conventional detectors) could provide an effective remedy. Appropriate 
selection could be implemented with a polarization-independent quantum
non-demolition measurement of the total photon number at Bob's end. 
Alternatively, pre-selection could be implemented by enhancing the
coupling between modes (2, 3) relative to modes (1, 4).

Despite our comments, the experiment of Bouwmeester {\it et al}.\ is a
significant achievement in demonstrating the non-local structure of
teleportation.

\end{document}